\RequirePackage{fix-cm}
\documentclass[twocolumn]{svjour3}          
\smartqed  
\usepackage{graphicx}
\usepackage{subcaption}
\usepackage{caption}
\usepackage{threeparttable}
\usepackage{booktabs}
\usepackage[dvipsnames]{xcolor}
\RequirePackage{hyperref}

\def\makeheadbox{\relax}

\begin{document}

\title{Security Analysis of Distributed Ledgers and Blockchains through Agent-based Simulation\footnotemark}

\author{Luca Serena \and
        Gabriele D'Angelo \and 
        Stefano Ferretti
}


\institute{G. D'Angelo \at
              University of Bologna (Italy) \\
              \email{g.dangelo@unibo.it}
}

\maketitle

\footnotetext{
\textbf{{\color{red} This is the authors' version of the following
article: ``Luca Serena, Gabriele D'Angelo, Stefano Ferretti. Security Analysis of Distributed Ledgers and Blockchains through Agent-based Simulation. To appear in Simulation Modelling Practice and Theory (Elsevier)''.}}}

\begin{abstract}
In this paper\footnote{An early version of this work appeared in \cite{gda-asia-2019}. This paper is an extensively revised and
extended version where more than 50\% is new material.}, we describe LUNES-Blockchain, an agent-based simulator of blockchains that relies on Parallel and Distributed Simulation (PADS) techniques to obtain high scalability. The software is organized as a multi-level simulator that permits to simulate a virtual environment, made of many nodes running the protocol of a specific Distributed Ledger Technology (DLT), such as the Bitcoin or the Ethereum blockchains. This virtual environment is executed on top of a lower-level Peer-to-Peer (P2P) network overlay, which can be structured based on different topologies and with a given number of nodes and edges. Functionalities at different levels of abstraction are managed separately, by different software modules and with different time granularity. This allows for accurate simulations, where (and when) it is needed, and enhances the simulation performance. Using LUNES-Blockchain, it is possible to simulate different types of attacks on the DLT. In this paper, we specifically focus on the P2P layer, considering the selfish mining, the 51\% attack and the Sybil attack. For which concerns selfish mining and the 51\% attack, our aim is to understand how much the hash-rate (i.e.~a general measure of the processing power in the blockchain network) of the attacker can influence the outcome of the misbehaviour. On the other hand, in the filtering denial of service (i.e.~Sybil Attack), we investigate which dissemination protocol in the underlying P2P network makes the system more resilient to a varying number of nodes that drop the messages. The results confirm the viability of the simulation-based techniques for the investigation of security aspects of DLTs.
\end{abstract}


\maketitle

\section{Introduction}
Distributed ledgers and blockchain-based technologies have become more and more popular over these years, due to their suitability to be used in many distributed application scenarios~\cite{D'Angelo:2018,cryblock2019}. Traceability, auditing, attestation as-a-service, pseudo-anonymity, and cooperation are just few examples, other than the traditional fintech applications that made these technologies famous.

A Distributed Ledger Technology (DLT) is a distributed software system that can be seen organized as a protocol stack. At its lower level, an overlay network is built on top of the Internet layer. This Peer-to-Peer (P2P) system is responsible to manage the dissemination of the messages. Each node is logically linked to a certain number of neighbors, and communicates with them based on a ``gossip-like'' dissemination protocol. Furthermore, some discovery mechanism is used, in order to build the P2P overlay~\cite{DBLP}. For instance, a random selection protocol is used in Bitcoin, while Ethereum employs a UDP-based node discovery mechanism inspired by Kademlia~\cite{10.1007/3-540-45748-8_5}.

The actual distributed ledgers are built on top of this kind of P2P system. Each node has its copy of the ledger that records all the transactions that have been generated in the network. Eventual consistency of the view of the distributed ledger is reached through the adoption of a consensus scheme. There are several protocols that may be adopted for this task. Usually, blockchains are based on ``Proof of Work'', ``Proof of Stake'' or one of their variants.

Finally, on top of the consensus layer we have the real transaction ledger, that records transactions and data. The implementation of the ledger varies, depending on the technology. In Bitcoin and Ethereum, for instance, the ledger is organized as a sequence of blocks, each block containing a set of transactions (for this reason, they are called ``blockchains''). Some newer alternative solutions adopt different data structures. Just as an example, the Tangle ledger employed in IOTA is in fact a directed acyclic graph of transactions. Moreover, in traditional blockchains, such as Bitcoin, the transactions that are inserted into the ledger just contain exchanges of cryptocurrencies. Instead, the so called blockchains 2.0 (and further) allow for ``programmable transactions'', i.e.~the system is able to insert into the blockchain smart contracts, actual contracts written with a Turing complete scripting language~\cite{D'Angelo:2018}. Such smart contracts can be triggered through specific transactions. Once activated, a related code is executed whose output changes the internal state of the smart contract.

While a layered organization of a decentralized system, such as DLTs, allows us to isolate the very different aspects of the technology and to obtain a better understanding of the components' protocols, the functioning of each layer influences the performance of the other layers. Thus, it becomes interesting to evaluate all possible alternatives of each component, and how possible modifications affect other aspects of the DLT. However, the complexity of DLTs, and the large scale nature of this distributed system make extremely difficult such evaluation process. This is especially true when dealing with security issues. For this reason, the creation of a virtual environment, in which DLTs are simulated, might represent a viable and profitable strategy.

In this work, we present a virtual environment called LUNES-Blockchain. The main component of LUNES-Blockchain is a discrete event simulator that is able to simulate the blockchain behaviour and to trigger certain attacks on such a decentralized data structure. The software has been designed to be scalable, thus it follows a Parallel And Distributed Simulation (PADS) approach. More in detail, the simulator allows us to create a synthetic DLT, organized as a virtual environment of different interacting agents. It is worth noticing that LUNES-blockchain has been designed and implemented to follow a multilevel approach. In fact, each level of the virtual environment simulates a different layer of a DLT protocol stack. Thus, a P2P level has been implemented, which is in charge of the modelling the overlay creation, management and message dissemination. The upper level models the specific DLT consensus scheme and the internals needed to maintain a consistent version of the ledger. (For example, as concerns Bitcoin, the system simulates the Proof-of-Work -- PoW mining approach, in which each node in the blockchain concurrently tries to mine a novel block.)

The adoption of a multi-level approach has the advantage that each level can be configured to work with a specific time granularity, based on the specific nature of the events that need to be simulated at different layers. For instance, the events generated at the (upper) blockchain level can be represented at a coarser grain than the (underlying) overlay level. In fact, the generation of a novel transaction, or a block in the blockchain corresponds to multiple dissemination steps at the overlay level, where messages are broadcast through the overlay, i.e.~they are transmitted from a node to its neighbors, and so on. Following this approach, a virtual environment composed of DLT nodes is created and it is implemented by a further lower-level simulation middleware as an agent-based framework on top of a PADS system.

The resulting virtual environment is able to model the behavior of a blockchain with a high degree of fidelity and to obtain fast results even in presence of complex or large-size networks to model. LUNES-Blockchain can be used as a testbed to study new components to be integrated in the DLT, to evaluate the performance of existing algorithms or to verify the reliability of the blockchain in presence of failures or malicious actors. In this paper, the developed virtual environment is used to study some well-known security attacks to the Bitcoin blockchain, i.e.~i) a Denial-of-Service (DoS) attack (also known as Sybil attack) \cite{Eyal:2018:MEB:3234519.3212998}, ii) selfish mining \cite{selfish}, and iii) 51\% attack \cite{51}. Indeed, we claim that the use of simulation techniques as a part of highly realistic virtual environments can be a profitable approach to study and assess the security aspects of a system.

As concerns the evaluation reported in this paper, in order to show the feasibility of the virtual environment based on simulation, we provide results related to the execution of the mentioned security attacks when applied to different configurations both at the DLT level and at the overlay level. For instances by changing the overlay topology and the specific configuration of the gossip dissemination strategy. We report on both the outcomes of the attacks, thus providing results related to the specific simulated virtual environment, as well as the performance of the simulator itself, in terms of time necessary to complete the simulation execution and scalability. The obtained results demonstrate the viability of exploiting agent-based simulation for building virtual environments of DLTs.

The remainder of this paper is organized as follows. Section \ref{sec:back} provides some background and a discussion on related works. Section \ref{sec:simulator} describes the virtual environment. Section \ref{sec:perf} is related to the presentation of the security attacks and to the usage of the virtual environment to assess their impact on the Bitcoin blockchain. Section \ref{sec:simresults} provides experimental results, with respect to the performance evaluation about the simulation tool we built. Finally, Section \ref{sec:conc} provides some conclusions and final remarks.

\section{Background and Related Work}\label{sec:back}
The blockchain is a technology that was initially proposed in the Bitcoin system, in 2009, by an anonymous author with the pseudonym of Satoshi Nakamoto~\cite{Nakamoto_bitcoin}. It can be seen as a decentralized and immutable database where all participants can transact directly without relying on an intermediary, such as a bank. Crucial for the proper functioning of the system is the role of cryptography, as well as the use of a consensus protocol, which ensures that all the various nodes agree on the content of the distributed ledger. Usually the users are identified with a pseudonym derived from their cryptographic public key, so while transactions are known in the ledger, it is difficult to identify the involved parties.

While Bitcoin is still currently the most famous and the most commonly used blockchain-based system, certain kinds of platforms, like Ethereum, are gaining popularity because they enable to develop more complex applications through the use of Smart Contracts, which are actual contracts written with a Turing complete scripting language. In fact, cryptocurrencies like Bitcoin basically only allow for money transfer, while blockchains based on smart contracts make it possible to create distributed applications. The execution of the public code composing the smart contract is carried out by the multiple nodes that are part of the system.

\subsection{The Peer-to-Peer Overlay and the Distributed Ledger}
\label{gossip-protocols}
The blockchain is a type of distributed ledger, where information is grouped into blocks, which are logically linked among each other through the use of a hash pointer to a previous block. Thus, blocks are organized in a chain, and forks can occur. In the case of forks, usually the strategy is to continue the longest chain, ignoring shorter ones.
At the distributed system layer, nodes are organized as a peer-to-peer network, in which each active participant has a copy of the shared ledger. For the communication among the nodes the underlying network is exploited, using the typical Internet protocols.

Each novel transaction, generated by a node, is disseminated through the P2P system using a flooding dissemination protocol~\cite{gda-jpdc-2017,Ferretti20131631}. More specifically, these protocols (which are usually referred as ``gossip protocols'') can implement different strategies to maximize the dissemination of the messages while controlling the communication overhead. The simplest approach, for a node that is part of the network, is to broadcast the newly received message to all the neighbors except the forwarder. A caching mechanism and a time-to-live associated with the messages will prevent infinite loops of messages. However, there are more complex strategies that enable us to minimize the amount of messages sent or to enhance the anonymity properties of the sender of a specific message (i.e.~a transaction recorded in the blockchain). In the following we briefly introduce the most common dissemination strategies.
\begin{itemize}
\item \emph{Fixed Probability}: a message is sent to a neighbor only if a random-generated number is greater than a certain threshold value (that is a parameter that can be tuned). The operation is then repeated for every neighbor of the node, except for the forwarder of the message.
\item \emph{Probabilistic Broadcast}: a message is sent to all the neighbors (except the forwarder) only if a random-generated number is greater than a certain threshold value. Otherwise no message is sent.
\item \emph{Dandelion}: this dissemination protocol has the specific goal to enhance the anonymity of the sender of a transaction. It consists in two main phases: in the initial ``stem phase'', the message is forwarded to a single random neighbor. In the following, the information is broadcast into the network in what is called the ``fluff phase''.
\item \emph{Dandelion+-}: based on Dandelion++ ~\cite{10.1145/3224424}, it is an improved version of Dandelion that aims to strengthen the resilience against the de-anonymization of blockchain users and the execution of Sybil attacks. In particular, to avoid having lost transactions (caused by malicious or defective nodes that drop the messages during the stem phase), the protocol implements a fail-safe mechanism. In fact, if a node receives a message during the stem phase and it does not get it back after a certain amount of time in the fluff period, then such a node will start the fluff phase by itself, by broadcasting the message. In the actual implementation of Dandelion++ other changes with respect to Dandelion are performed. Particularly, in Dandelion++ nodes in one epoch (i.e.~a period of about $10$ minutes) are either ``relayers'' or ``diffusers'', and thus they relay the messages to either one neighbor or to all the neighbors. In the tests reported in this paper, for simplicity and for being able to evaluate the specific impact of the recovery mechanism, we decided to only consider the fail-safe mechanism among the upgrades of Dandelion++ (and we refer as ``Dandelion+-'' for this specific variant of the protocol.
\end{itemize}

\subsection{Consensus Scheme}
The consensus scheme is a protocol that ensures that all the nodes of the P2P network maintain the same view on the current state of the blockchain. The scheme defines what has to be done to validate a block, finding a common agreement among all the nodes of the network. In the past years, several consensus schemes have been proposed in the distributed systems research area. Some of these schemes are utilized today in blockchain technologies (e.g.~Practical Byzantine Fault Tolerance~\cite{Castro:1999:PBF:296806.296824}). However, the main approaches used in blockchain are the Proof-of-Work (Pow, used in Bitcoin and the actual Ethereum) and Proof-of-Stake (PoS, that is planned to be used in the novel Ethereum version)~\cite{DBLP}.

In PoW systems, in order to validate a block, nodes have to solve a computational crypto-puzzle that requires a very large amount of computational power. This activity is called mining and the users participating in the mining activity are called miners. More specifically, a valid block is found when one finds a specific number (called nonce) that, when hashed together with the proposed block, provides as a hash output a sufficiently small number, contained in a certain range. The difficulty of resolving the puzzle can be dynamically updated in order to adapt to the overall growth of computing power. When using the PoW, it may happen that more than one node mines a block at the same time, thus creating a fork in the chain of the mined blocks. When this occurs, the nodes agree to consider as valid only the blocks in the longest chain that will therefore be continued in the future. Conversely, the content of the shorter chains will be ignored. Proof-of-Work has often been criticized due to the enormous waste of energy caused by the mining activity.

The systems based on PoS are much less computation-consuming. The users who want to participate in the validation of a block must deposit a certain amount of cryptocurrency. The probability that a node is chosen as the next validator is proportional to the amount of currency deposited in the escrow. To reduce the likelihood of some kinds of attack, there may also be some mechanisms that prevent the same nodes from validating blocks too often.

Regardless of the protocol used, the hash of the previous block in the chain is inserted in the novel block, therefore making it easier to verify the ledger state in a tamper-proof manner. It is worth reminding that the work of validating a block is financially rewarded. The reward can be achieved by the act itself of validating a block and from the fees that the users pay to get their transactions validated.

\subsection{Simulation of Blockchains}
At the time of writing, the literature on blockchain simulators is still not abundant. Usually, the main focus is on the analysis of the blockchain, the use of smart contracts and sometimes the security issues of blockchain systems. The typical approach is to develop smart contracts and test them using local blockchains. Remix, Metamask, Ganache, Multichain and the Ethereum test networks (e.g.~Ropsten, Rinkeby) are examples of environments though to write, compile and debug smart contracts. In accordance with the multi-layered vision of a blockchain we discussed in the previous section, a common approach is to simulate just a few aspects of a blockchain at a time.

The majority of scientific literature about the simulation of blockchains is interested in predicting the performance of blockchain systems or in modelling specific aspects such as the consensus algorithms.

For example, in \cite{7930225}, the authors propose to use modelling and simulation techniques to predict the latency of blockchain-based systems.

The authors of \cite{10.1007/978-3-319-94478-4_2} describe the usage of SimPy to measure some metrics from a simulated Blockchain system and compute a general score that is called ``quality of blockchain''.

The focus of \cite{8632560} is on the mining process. More in detail, the authors have implemented the simulation of the mining process based on queuing theory. They extracted from the Bitcoin real data, the parameters necessary to simulate the process by means of a M/M/n/L queuing system evaluated using JSIMgraph.

In \cite{simblock}, a blockchain network simulator is presented. This simulator follows an event-driven approach to model the neighbor nodes selection of the peer-to-peer overlay. As in LUNES-Blockchain, the mining activity is not simulated in detail, but a block generation is mimicked based on the computational capabilities of nodes.

VIBES is a blockchain simulator, thought for large modeling scale P2P networks~\cite{Stoykov:2017}. The rationale behind this simulator is to provide a blockchain simulator that is not confined to the Bitcoin protocol, trying to support large-scale simulations with thousands of nodes.

BlockSim is composed of a Python framework able to build discrete-event dynamic system models for blockchain systems~\cite{Alharby:2019}. BlockSim is organized in three layers: incentive layer, connector layer and system layer. Particular emphasis is given on the modeling and simulation of block creation through PoW.

In \cite{191667} is described a new methodology that enables the direct execution of multi-threaded applications inside of Shadow that is an existing parallel discrete-event network simulation framework. This is used to implement a new Shadow plug-in that directly executes the Bitcoin reference client software (i.e.~Shadow-Bitcoin).


A small number of research activities has focused on the usage of simulation-based techniques to evaluate the security of the current blockchain systems or to drive their future development.

The simulation-based evaluation of a very specific kind of blockchain, i.e.~Tangle-based blockchains, is discussed in \cite{10.1007/978-3-030-55304-3_35}. The paper shows how a simple simulation model can be defined to model the Parasite Attack, Double Spending Attacks and Hybrid Attacks on top of a Tangle.

The authors of \cite{8802494} propose to use an agent-based simulation to guide and optimize protocol development in a Proof-of-Stake parameter optimization and peer-to-peer networking design. In this work, they describe a system for simulating how adversarial agents, both economically rational and Byzantine, interact with a blockchain protocol.

In \cite{Gervais:2016}, the mining strategy of Bitcoin is simulated and studied. A network is modeled, but the propagation of transactions is not simulated, since the focal point is to study the impact of the block size, block interval, and the block request management system.


To the best of our knowledge, LUNES-Blockchain is the first simulator of blockchains that is able to take advantage of the performance speedup and extended scalability provided by PADS. 
A detailed performance evaluation of the speedup that can be obtained by employing PADS techniques in the simulation of blockchains is left as future work since in this paper we prefer to focus on the security aspects of DLTs that can be investigated by means of simulation. However, the simulation of blockchains has a lot common with the simulation of complex networks. For example, the communication as represented in the peer-to-peer network used to spread messages in the distributed system that builds up the DTL is not different from a message dissemination on a complex network (i.e.~gossiping on a scale-free network). On this side, PADS techniques have already demonstrated that they can significantly reduce the amount of time that is required to complete the simulation runs while enabling the simulation of larger network topologies~\cite{annsim2021,gda-jpdc-2017,10.1145/2486092.2486115,10.4108/ICST.SIMUTOOLS2009.5672,10.1109/DS-RT.2016.17}.
\subsection{Security Analysis Through Virtual Environments}
Virtual environments are an important tool for studying the security, efficiency and scalability of a computer system. Sometimes, depending on the complexity of the systems, it is not possible to analyze certain specific problems only by analytic methods or model checking, thus the creation of a virtual environment that relies on the usage of simulation techniques may be needed. In certain situations, it could be necessary to build a model characterized by high fidelity, maybe including in the virtual testbed also hardware components. Under other circumstances, on the other hand, it might not be necessary to represent all the system in its complexity but some aspects can be neglected, thus favoring the execution speed of the virtual environment and its scalability.

In particular, the creation of a virtual environment based on simulation has the advantage that made it possible to study and detect issues in advance with respect to the real implementation of the system to be built. For example, it is possible to carefully analyze critical points and possible vulnerabilities of the applications that are going to be developed. This can be useful above all where the actual implementation is costly or where potential bugs could lead to dire or not easily fixable consequences. Another well-known example are digital twins. A digital twin is defined as a digital replica of a living or non-living physical entity~\cite{twinsec}. That replica can be a very useful testbed for existing systems as well, since it allows us to test and evaluate strategies for solving or improving certain known issues without shutting down the actual system.

While most of the studies about cybersecurity focus on \textit{confidentiality} (i.e.~data can be accessed only by those who have the permission), \textit{integrity} (i.e.~data are not altered or corrupted) and \textit{availability} (i.e.~data and services are always available to those who have the permission whenever they want), this work aims to investigate the usage of virtual environments for the study of high level attacks, specifically to the overlay network (i.e.~Sybil attack) and to the consensus protocol (51\% and selfish mining attacks).

The focus of this paper is on attacks that, in some way, involve the whole network. We decided to study this kind of attacks because they are among the major risks in the current blockchains and because they represent a good opportunity to show the scalability of LUNES-Blockchain. On the other hand, it is worth to note that virtual environments can be also used to study targeted attacks on the virtualized systems or to investigate specific conditions of given nodes in the blockchain network. For example, we could be interested in evaluating the security aspects of certain dissemination protocols in presence of poorly connected nodes. In fact, the structuring of the overlay network can affect the level of coverage and efficiency that can be achieved in the system, with the least connected nodes behaving as most vulnerable ones. LUNES-blockchain allows us to easily test the network configurations that may present criticalities (i.e.~corner cases).

\section{Simulation of the Bitcoin Network}\label{sec:simulator}
With the aim to make this paper as self-contained as possible, in this section we introduce some background on Discrete Event Simulation (DES) and Parallel And Distributed Simulation (PADS) techniques. After that, we describe the ART\`IS/GAIA simulation middleware and the LUNES simulation model that have been used for implementing LUNES-Blockchain.

DES is a simulation paradigm that allows us to combine likelihood of the representation and ease of use. In DES, each event (i.e.~a change in the simulated model) happens at a specific time in the model evolution. In other words, a DES can be seen as the execution in chronological order of a sequence of events. Under the implementation viewpoint, the main components of a DES are i) a set of state variables that model to the various entities in the modelled system, ii) an ordered list of future events to be executed and iii) a global clock that represents the current time in the simulation and that triggers the execution of events from the ordered list.

A DES can be implemented in many different ways. For example, it is possible to rely on both monolithic or distributed architectures. In a monolithic (i.e.~sequential) simulation, all the model state variables representing the simulated model are allocated in a single execution process (usually run on top of a single CPU core) that is in charge of managing and generating events. The main advantage of this approach is its simplicity but often its execution speed is not optimal and it lacks of scalability. For example, the scalability of the simulator is limited, both in terms of time required to complete the simulation runs and complexity of the system that can be modelled~\cite{1668384}.

When it is necessary to boost execution performances or to model complex systems then it is possible to follow a PADS approach. In this case, the simulation is executed on top of a parallel/distributed execution architecture composed of multiple interconnected CPUs. A PADS permits to share the model workload among multiple LPs (i.e.~logical processes, model components executed by single execution units). Following this approach there is no more a global state of the simulation model but each LP manages a specific portion of the simulated model. In practice, each LP (that is now run on a different CPU core) deals with just a part of the pending events and the various LPs communicate via messages to deliver remote events. The main advantage of using a PADS is that it enables the modelling and the processing of larger and more complex simulation models with respect to a monolithic setup. On the other hand, the partitioning of the simulated model is not a complex problem~\cite{gda-simpat-2017} and the usage of a synchronization algorithm that coordinates the LPs execution is needed to guarantee the simulation correctness~\cite{FUJ00}. In other words, the output obtained by a PADS simulation must be exactly the same as what obtained when running the simulation model by using a sequential approach.

\subsection{ART\`IS/GAIA}
The \textit{Advanced RTI System} (ART\`IS) is a parallel and distributed simulation middleware that implements the partitioning of the simulation model in a set of LPs. Following this approach, the simulation is run on top of a parallel/distributed execution architecture that is composed of multiple interconnected Physical Execution Units (PEUs). Each PEU (i.e.~a CPU core) runs at least one LP. The simulation middleware (i.e.~ART\`IS) is in charge of providing to the LPs some main services like i) time management (i.e.~synchronization), that is necessary for obtaining correct simulation results and ii) data distribution management, that is necessary for the efficient delivery of interactions between the LPs.

The communication between the different partitions of the simulated model (i.e.~LPs) is very relevant in terms of overhead for the execution of a PADS. This is due to the amount of time that is spent in delivering the interactions between the model components. In other words, the amount of time that is necessary to complete the execution of a PADS (i.e.~wall-clock time) is highly dependent on the performance of the communication network (i.e.~latency, bandwidth and jitter) that interconnects the PEUs and how the partitioning of the simulation model has been arranged.

The \textit{Generic Adaptive Interaction Architecture} (GAIA) is a software framework that uses the service provided by ART\`IS~\cite{pads} and that has two main goals: to simplify the definition of PADS models and to increase the efficiency of their execution. To achieve these goals, the simulation model is partitioned in a set of Simulated Entities (SEs) that can be seen as fine-grained model components. Each LP clusters a set of SEs and provides to them some basic simulation services (e.g.~synchronization and message passing) and some higher level services (e.g.~proximity detection, advanced data structures etc.). Following the approach implemented in GAIA, the evolution of the simulation model is obtained through interactions among the SEs. These interactions are then encapsulated by timestamped messages that are exchanged between the LPs. From the simulation modelling viewpoint, GAIA follows a Multi Agent System (MAS) approach in which each SE represents an agent. Depending on the simulation model that is implemented using GAIA, the SE can be a network node, a car or any other object used by the simulation model. More generally, each SE can be seen as an autonomous agent able to perform some specific actions (i.e.~individual behavior) and to interact with other agents in the simulation (i.e.~group behaviors).

In addition, GAIA implements a rather complex set of mechanisms that are aimed to reduce the communication overhead in the parallel/distributed execution setup. This is obtained by clustering the SEs that have frequent interactions in the same LP. For example, if a group of SEs is close in the simulated space, and they communicate using proximity-based technology (e.g.~WiFi networks), then it is likely a good idea to cluster them together. In terms of communication overhead, clustering the heavily-interacting entities has the advantage to reduce the amount of costly LAN/WAN/Internet communications while increasing the usage of efficient shared memory messages. In the current version of GAIA, this is obtained by analyzing the interaction pattern of each SE during the simulation execution and then triggering the necessary SEs re-allocations (i.e.~migrations) between the LPs. The decision on when and how to perform migrations can be taken in many different ways. Currently, GAIA implements the clustering of SEs using a set of high-level heuristics that analyze the communication behavior and that are independent form the specific simulation domain.

\subsection{LUNES}
\begin{figure}[t]
  \centering
    \includegraphics[width=0.5\textwidth]{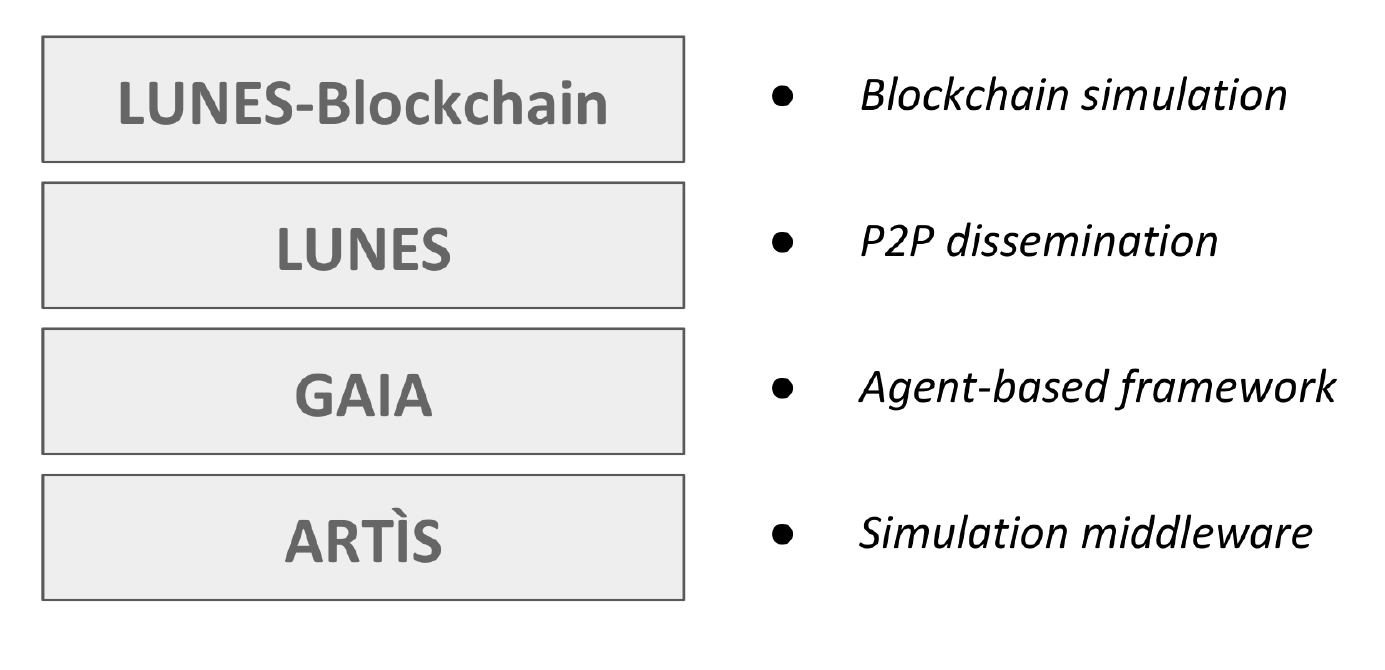}
  \caption{LUNES-Blockchain system architecture: the blockchain simulator (i.e.~LUNES-Blockchain) runs on top of the (simulated) communication services provided by the P2P dissemination simulator (i.e.~LUNES).}
  \label{fig:lunes-structure}
\end{figure}
LUNES (Large Unstructured NEtwork Simulator) is a simulator of complex networks implemented on top of ART\`IS/GAIA. The software has been designed to simulate large scale peer-to-peer networks and to evaluate the efficiency of data dissemination protocols running on top of them. The tool is thought to be easily expandable \cite{gda-jpdc-2017}, so that adding new simulation models on top of the existing services is a rather simple task.

The simulator has a modular structure in which each module is in charge of a specific phase that is executed separately:
\begin{itemize}
    \item \emph{Creation of network topology}, which is entrusted to an external library called  ``igraph``~\cite{igraph}. There is the possibility to choose among different network topologies, as well as setting the network size (i.e.~number of nodes) and the total number of edges.
    \item \emph{Protocol simulation}, that implements the data dissemination on top of the peer-to-peer network for a given amount of simulated time. If there is a simulation model implemented on top of the dissemination protocol, then it is executed in this phase.
    \item \emph{Performance evaluation}, batch data evaluation and computation of main metrics on the data collected during the simulation.
\end{itemize}
Given that, LUNES is built on top of ART\`IS (i.e.~parallel and distributed processing) and GAIA (i.e.~adaptive self-clustering, dynamic computational and communication load-balancing), it can benefit of an efficient approach to simulate detailed protocols on top of large scale networks. For example, some tests on the hardware used for the performance evaluation reported in the following of this paper, showed that running LUNES using 4 LPs (on top of 4 different PEUs) permits to almost halve the time required to complete the simulation runs with respect to a sequential simulation (i.e.~using a single LP).

\subsection{LUNES-Blockchain}
LUNES-Blockchain is a virtual environment based on LUNES, whose structure is described in Figure~\ref{fig:lunes-structure}. This virtual environment allows us to study both the behaviour of the blockchain and the main security attacks on the modelled system. In LUNES-blockchain, each node is represented by means of an agent that implements a local behavior and interacts with other agents. This kind of representation simplifies the development of the model and, in our view, it adds a high-level of extensibility to the virtual environment.

LUNES-Blockchain specifically simulates the behaviour of a blockchain based on PoW, such as Bitcoin. Some of the nodes are defined to be miners, and it is assigned to them a specific hash-rate, corresponding to the percentage of computational power at the disposal of those nodes. Furthermore, the software allows us to assign to certain nodes, representing computationally powerful clusters, a bigger amount of cryptographic power. Finally, during the simulation of the attacks, a specific hash-rate is assigned to the malicious node, thus scaling back the hash-rates of the other nodes.

In the virtual environment, the time in the simulated systems is represented  by using a sequence of time-steps. As mentioned before, to increase the veracity of the simulation, the virtual environment uses a multilevel approach, in order to represent separately the mining steps (about $5$ minutes long) and the propagation steps (a few fractions of a second). In LUNES-Blockchain, the simulation of the mining process is modelled to follow the difficulty and the behavior of the Bitcoin mining process, where a block is mined on average every $10$ minutes. This multi-level approach permits to model the fine details when it is necessary without affecting the simulator scalability.

The design of LUNES-Blockchain has been structured in phases: i) modelling and simulation of a generic blockchain; ii) modelling of the specific aspects of the Bitcoin blockchain; iii) modelling of some specific attacks to the blockchain (i.e.~``51\% attack'', ``denial of service'' and ``selfish mining''). This multi-phase approach, permits to easily add the missing functionalities (e.g.~another type of attack) or to add the support for another blockchains.

According to the Bitcoin policy, the Bitcoin simulation model included in LUNES-Blockchain (i.e.~phase ii), propagates the new blocks using a broadcast dissemination mechanism \cite{decker2013information}. Despite that, LUNES-Blockchain allows the use of different dissemination protocols as well. This enables to evaluate the impact of these protocols on the blockchain performance and their effects on the blockchain security. For example, when evaluating the denial of service attack, the different dissemination protocols provided by LUNES led to different spreading of messages on the P2P network. This permits the use of the LUNES-Blockchain virtual environment to investigate the resilience of the dissemination algorithms, as well as of consensus protocols, against a certain number of nodes that do not relay the received messages.
LUNES-blockchain is a high level simulator, which aims to reproduce only the relevant aspects for our purposes, thus lowering the complexity and enhancing the ease of management. For example, Proof-of-Work is replaced by the probability for a node to mine a block in a certain time interval and the blocks in the simulations are not structured with all the effective components, because reproducing an actual exchange of transactions would be meaningless for our results and would only bring a significant time overhead in the execution. In accordance to this principle, in the simulations of the Sybil attack the behaviour of the blockchain is not reproduced, since this attack is not specific of the crypto-systems, but it can be applicable to all the peer-to-peer systems based on the relays of messages.
Finally, the software is thought to allow the users to easily modify execution parameters such as the employed gossip protocol, the overall number of time-steps, the percentage of nodes acting as miners and the difficulty for a node to mine a block. Regarding the latter, in our default configuration a block is mined on average every two mining time-steps, meaning that each time-step represents a time slot of about 5 minutes.

LUNES-Blockchain is available for peer-review and it will be included in the forthcoming release of LUNES that will be available in source code format on the research group website~\cite{pads}.

\section{Evaluation of Security Attacks}\label{sec:perf}

In this section, we report the results obtained using LUNES-Blockchain to study some well-known security attacks on the simulated Bitcoin blockchain. In particular, regarding to the 51\% attack, we study the relationship between the hash-rate of the attacker and his/her ability to control the flow of the generation of blocks, which could lead to a misbehaviour like the double spending~\cite{rosenfeld2014analysis}. About selfish mining, we evaluate the feasibility of the attack and the actual benefits for the attacker of such a strategy. Finally, for which concerns the Sybil attack, we analyze the role of the dissemination protocol used in the underlying peer-to-peer network (i.e.~gossip protocol) and the topology of the network in resisting to the attempt to isolate specific nodes from the system.

\subsection{51\% Attack}
51\% attacks occur when a node, or a group of colluded nodes, owns a significant part (if not the majority) of the hashing power of the blockchain network, thus undermining the decentralized nature of the system. Attackers can exploit their computational power to mine a high amount of blocks in order to carry out fraudulent behaviour, such as double spending.
It is important to notice that, in order to have a majority of the hash-rate and be sure to have enough power to control the evolution of the blockchain, the attacker actually needs a hash-rate strictly higher than 50\% (i.e.~$50\%+1$). However, for the sake of a simpler name, people usually refer to this situation as ``51\%". Thus, we adopt the same terminology.

Three different criteria have been used to evaluate how much the cryptographic power, owned by the attacker, can influence its ability to control the flow of blocks' generation. The metrics used are:
\begin{itemize}
    \item Number of blocks actually mined by the attacker and then inserted in the main chain;
    \item Percentage of blocks in the main chain mined by the attacker;
    \item Percentage of blocks mined by the attacker which ends up in the main chain.
\end{itemize}
The LUNES-Blockchain virtual environment has been configured to simulate a random graph composed of $500$ nodes and $2000$ undirected edges. In this configuration of the simulator, $99$ tests have been made, each one with an increasing percentage of the attacker's hash-rate (from $1\%$ to $99\%$). For each of the $99$ tests several runs have been executed and the average result have been calculated and reported. Every setup has been evaluated twice, in order to check if the presence of the pools of miners (i.e.~miners who group together sharing their cryptographic power to have a better chance to mine a block) can influence the attack outcomes. In the first configuration, we simulated the presence of $9$ nodes, representing the biggest known Bitcoin pools and constituting all together the $80.4\%$ of the hash-rate of the network (attacker excluded). In the second configuration, pools are not considered, meaning that there are no nodes with a hash-rate significantly higher than the others.

\begin{figure}[t]
  \centering
    \includegraphics[width=0.5\textwidth]{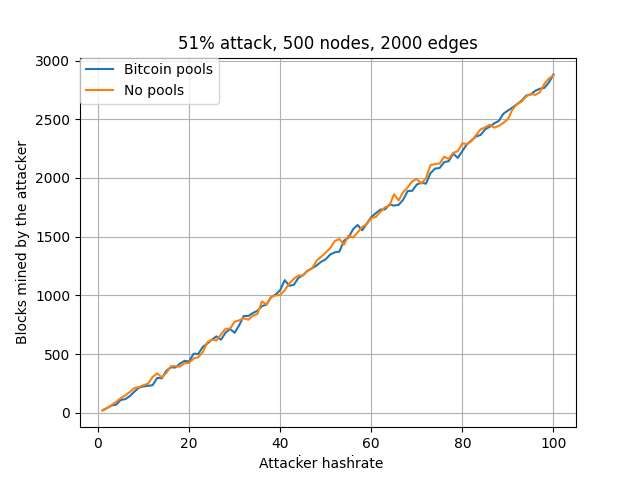}
  \caption{The X-axis indicates the percentage of cryptographic power owned by the attacker while the Y-axis indicates the number of blocks mined by the attacker.}
  \label{51totalmined}
\end{figure}

{The results show that, unsurprisingly, the higher the hash-rate of the attacker the higher the number of blocks mined by the attacker, since the ability of mining a block for a node is proportional to its computational power. The curve in Figure~\ref{51totalmined} has an almost linear behaviour. It is worth noticing that, on this aspect, the presence of mining pools does not affect the behavior of the attack.

Figure~\ref{percentageAttacker} shows a relevant difference between the expected and the observed behaviour of the attacker (i.e.~it was expected for a node owning $h$\% of the hash-rate to validate approximately $h$\% of the blocks in the main chain). With the attacker having a hash-rate greater than 25\%, the percentage of the blocks mined by the malicious nodes is always greater than what is expected, with a discrepancy around 10\% for certain values. The motivation of such a behaviour is that when a fork occurs, the nodes are encouraged to extend the part of the forked chain that contains their own validated blocks. LUNES-Blockchain implements such a strategy. Figure~\ref{percentageAttackerTotal} shows that the percentage of the blocks mined by the attacker among all the blocks (and not just the ones that end up in the main chain) follows the expected behaviour. This confirms that the discrepancy noted in Figure~\ref{percentageAttacker} occurs just because of the choice of which chain to extend.
Furthermore, from figure \ref{percentageAttacker} one can deduce the probability to perform double spending, since in order to accomplish such fraudulent behaviour the attacker needs to mine the next block that is going to appear in the main chain. Our results agree with studies like \cite{51}, stating that for attackers with medium-low hash-rate it is difficult (but not impossible) to succeed in double spending, while for attackers with a high hash-rate the success of the attack is much more probable.

\begin{figure}[t]
  \centering
    \includegraphics[width=0.5\textwidth]{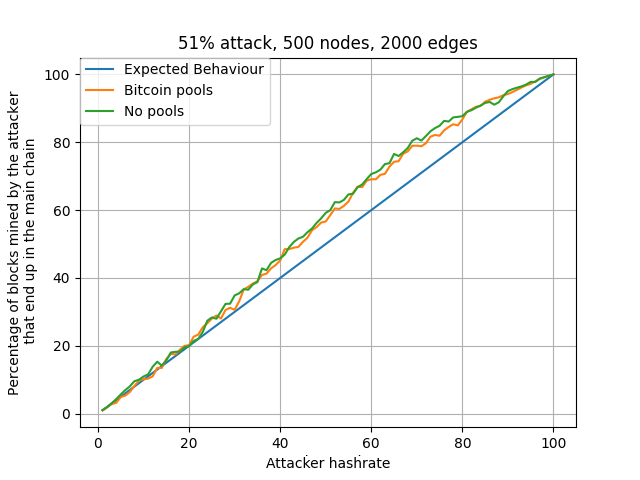}
  \caption{The X-axis indicates the percentage of the cryptographic power owned by the attacker while the Y-axis indicates the percentage of blocks in the main chain mined by the attacker. The blue line indicates the expected behaviour while the orange curve reports the observed behaviour.}
  \label{percentageAttacker}
\end{figure}

\begin{figure}[t]
  \centering
    \includegraphics[width=0.5\textwidth]{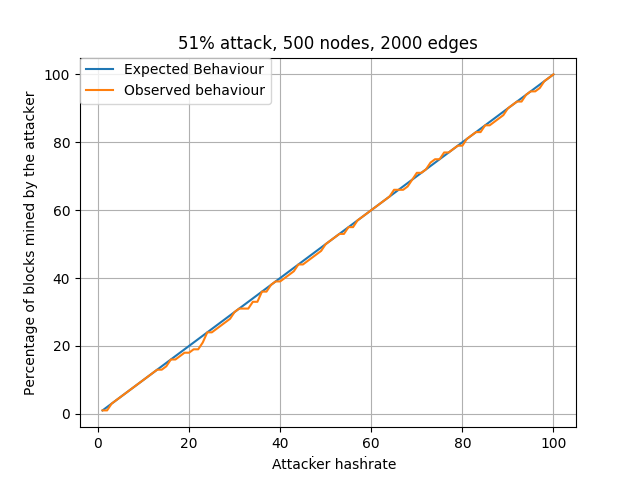}
  \caption{The X-axis indicates the percentage of the cryptographic power owned by the attacker while the Y-axis indicates the percentage of the total blocks mined by the attacker. The blue line indicates the expected behaviour while the orange curve reports the observed behaviour.}
  \label{percentageAttackerTotal}
\end{figure}

Figure~\ref{51goesmain} shows the percentage of the blocks mined by the attacker that actually end up in the main chain. The observed curve is more irregular if compared to the results presented above. However, the trend that, the higher the attacker's hash-rate the higher the percentage of blocks ending up in the main chain, is still evident. This happens for two main reasons: i) with a growing hash-rate owned by the attacker, there are less blocks that are mined by other nodes; ii) in case of forks, the malicious user will continue to extend the chain where there is a larger amount of blocks mined by him.

\begin{figure}[t]
  \centering
    \includegraphics[width=0.5\textwidth]{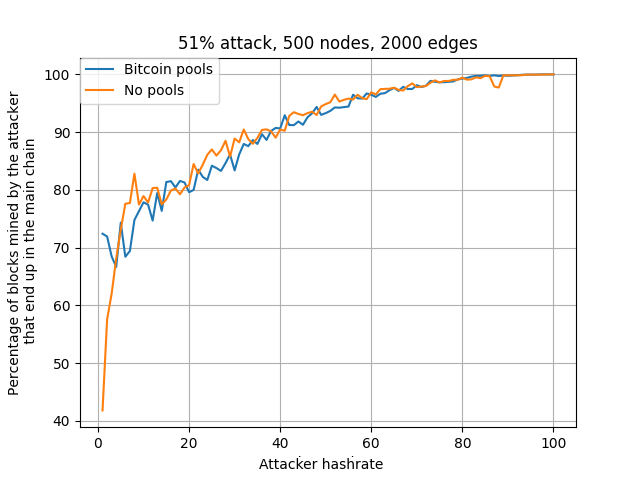}
  \caption{The X-axis indicates the percentage of cryptographic power owned by the attacker while the Y-axis indicates the percentage of blocks mined by the attacker that actually ended up in the main chain.}
  \label{51goesmain}
\end{figure}

It is relevant to note that the results reported in this section show that the presence of mining pools does not significantly affect the outcome of 51\% attacks.

\subsection{Selfish Mining Attack}
Selfish mining is an attack in which a malicious node does not spread immediately to the network the last block that it has mined. In fact, it tries to mine further blocks with the aim to gain an advantage of at least two unities with respect to the public ledger. In case of success, the computational effort of the other nodes becomes useless. This happens because, in the case of forks, the nodes agree to consider as valid and extend the longest chain. Thus, if the malicious node is able to locally mine a sequence of blocks longer than the main chain of the public blockchain, then the computational effort of the other miners will be wasted until when the attacker reveals its blocks. Using the LUNES-Blockchain virtual environment, we have tested how the blockchain behaves when an attacker has an advantage of $2$ or $3$ units before spreading its blocks.

Similarly to the 51\% attack, the selfish mining attack was investigated in the presence of a random graph with $500$ nodes and $2000$ edges. Also in this case, we varied the attacker hash-rate in order to better figure out the outcomes of the attack in different conditions.

\begin{figure}[t]
  \centering
    \includegraphics[width=0.5\textwidth]{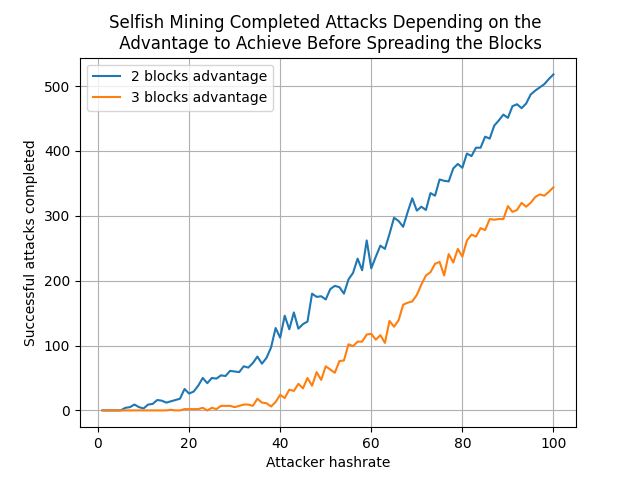}
  \caption{Selfish Mining Attack executed on a simulated network with random graph topology, $500$ nodes and $2000$ edges.}
  \label{selfish}
\end{figure}

Figure~\ref{selfish} shows that, as expected, increasing the hash-rate owned by the attacker leads to a greater number of finalized attacks. It is worth noticing that our tests demonstrate that selfish mining is not always convenient for the attacker. In fact, if an attacker that owns $50\%$ or less of the total hash-rate applies a selfish mining strategy, then the attacker is able to spread in the public main chain fewer blocks than when he/she implements a normal behaviour (i.e.~not malicious). Conversely, when the attacker owns a very high hash-rate (i.e.~more than a half of the overall computational power) then this attack strategy can be convenient. Figure~\ref{selfish-main-chain} reports the measured efficiency of the selfish mining strategy.

\begin{figure}[t]
  \centering
    \includegraphics[width=0.5\textwidth]{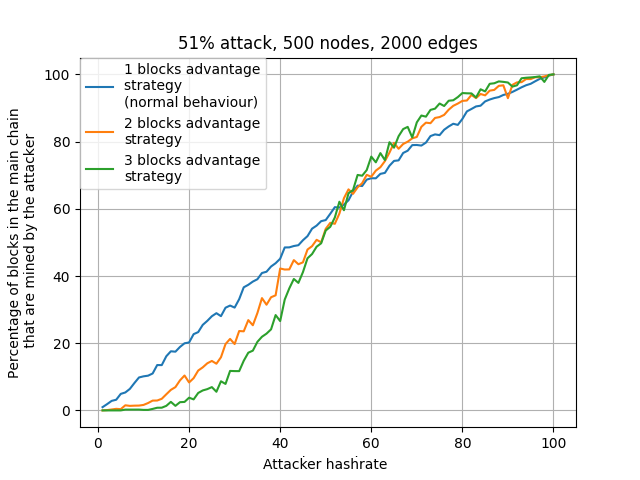}
  \caption{Number of attacker's blocks that ended up in the main chain. This number indicates the advantage (in terms of blocks) that the attacker wants to achieve before spreading to the world the blocks mined locally (and not broadcasted to the whole network yet).}
  \label{selfish-main-chain}
\end{figure}

\subsection{Sybil Attack}
A Sybil Attack is a specific type of denial of service in which a malicious user (or a group of colluded users) tries to gain a disproportionate influence in the network by creating a large number of fake identities directly controlled by him/her. More specifically, in this case the goal of the attacker is to isolate some specific nodes from the system, impeding them to spread their transactions to the network. Thus, when a Sybil node (~i.e. a node controlled by the attacker) receives a message created by the victim, it does not forward such a message to its neighbors. On the other hand, Sybil nodes behave according to the dissemination protocol for all the other messages received.

In the tests described in this section, we report the outcomes obtained from the LUNES-Blockchain virtual environment when the attack is executed on random graphs composed of $10000$ nodes. In each simulation run, we collected the percentage of potential receivers (i.e.~the nodes of the system except the attackers and the creator of the transaction) that are actually able to receive the newly generated message that has been delivered to the network. As described above, the goal of the attackers is to limit the diffusion of messages (i.e.~containing the new transactions) up to the point to exclude some nodes from the updates. In order to obtain results that are not influenced by the position of the nodes (whether if involved in the creation of the message or in the attack) within the network, the tests were repeated several times, choosing each time a different node as a message originator and then selecting the malicious nodes randomly. In these experiments, unlike other analyzed attacks, the computational power owned by the nodes is irrelevant for the purpose of the simulation, thus no hash-rate is assigned to the nodes in order to lower the complexity of the software. The employed dissemination protocol and the percentage of malicious nodes in the network are the only variable elements considered in this analysis. 

We have firstly investigated a setup in which each network generated for modelling the message dissemination is composed of an average of $8$ edges per node~\cite{gda-asia-2019}. The aim of this evaluation is to show which dissemination protocols are able to ensure a greater level of resilience against a varying percentage of malicious nodes (i.e.~nodes that do not relay the received message, trying to isolate the victim user from the system).

Figure~\ref{random-coverage-dandelion} shows the results obtained by Dandelion (in two different configurations of the stem phase) and Dandelion+- (that is a variant of Dandelion++ described in Section~\ref{gossip-protocols}). Figures~\ref{random-coverage-broadcast} and \ref{random-coverage-fixed}, report the results obtained using the Probabilistic Broadcast and the Fixed Probability dissemination protocols. Also in this case, different setups for the dissemination protocols have been investigated (i.e.~dissemination probabilities). The outcomes show that Dandelion is extremely vulnerable to Sybil attacks if compared to the other algorithms. This situation can be significantly improved by switching to the Dandelion+- dissemination protocol. 
In this way, it is possible to have both better guarantees for the anonymity of the blockchain users and a good resilience against Sybil attacks. The Fixed Probability and the Probabilistic Broadcast protocols show similar results, but it is worth noticing that Fixed Probability constantly behaves slightly better.

\begin{figure}[t]
  \centering
    \includegraphics[width=0.5\textwidth]{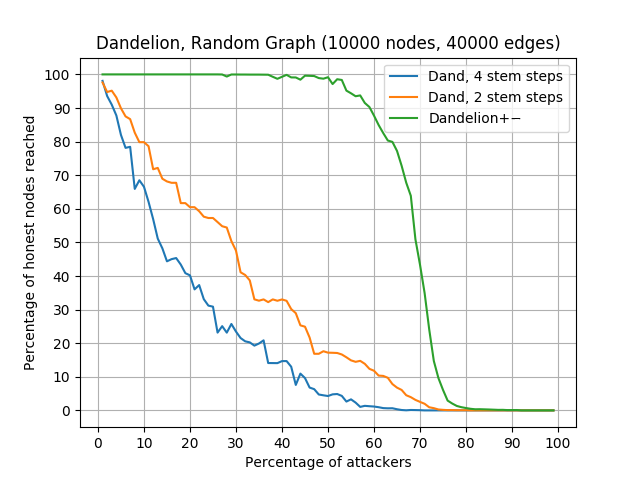}
  \caption{Coverage (Percentage of honest nodes reached) -- Dissemination: Dandelion, Overlay Topology: Random Graph with 40000 edges.}
  \label{random-coverage-dandelion}
\end{figure}
\begin{figure}[t]
 \centering
\includegraphics[width=0.5\textwidth]{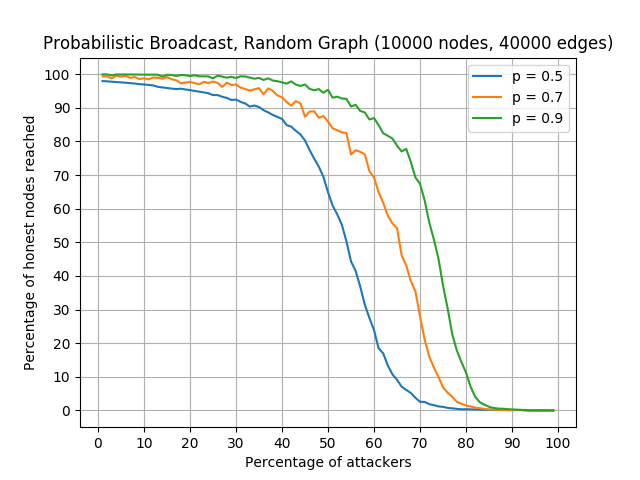}
  \caption{Coverage (Percentage of honest nodes reached) -- Dissemination: Probabilistic Broadcast, Overlay Topology: Random Graph with 40000 edges.}
  \label{random-coverage-broadcast}
\end{figure}
\begin{figure}[t]
  \centering
\includegraphics[width=0.5\textwidth]{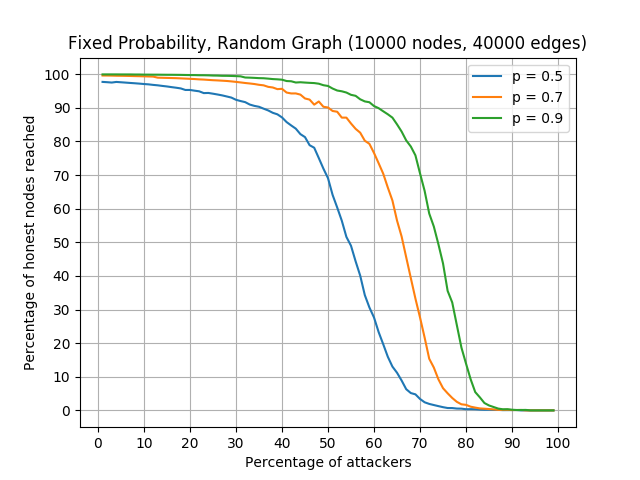}\\
  \caption{Coverage (Percentage of honest nodes reached) -- Dissemination: Fixed Probability, Overlay Topology: Random Graph with 40000 edges.}
  \label{random-coverage-fixed}
\end{figure}

Figures~\ref{random-coverage-dandelion-80000}, \ref{random-coverage-probabilistic-80000} and \ref{random-coverage-fixed-80000} report on the results obtained when doubling the number of edges in the network. As shown in the figures, in this case the attack is much more difficult to carry out, thus meaning that the average degree of the nodes is a key factor for withstanding the Sybil attack.

\begin{figure}[t]
  \centering
\includegraphics[width=0.5\textwidth]{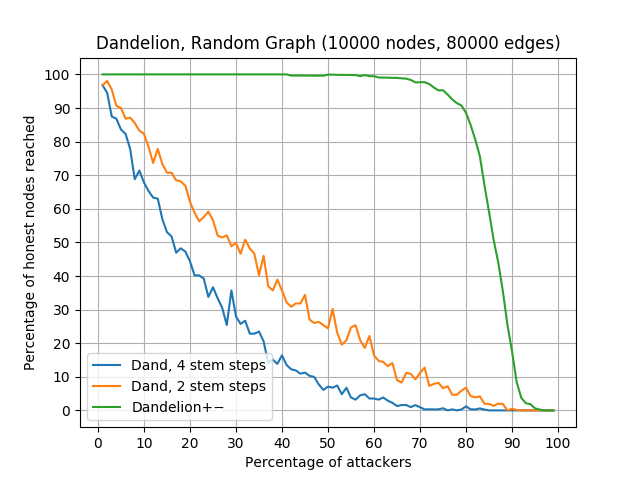}
  \caption{Coverage (Percentage of honest nodes reached) -- Dissemination: Dandelion, Overlay Topology: Random Graph with 80000 edges.}
  \label{random-coverage-dandelion-80000}
\end{figure}
\begin{figure}[t]
  \centering
\includegraphics[width=0.5\textwidth]{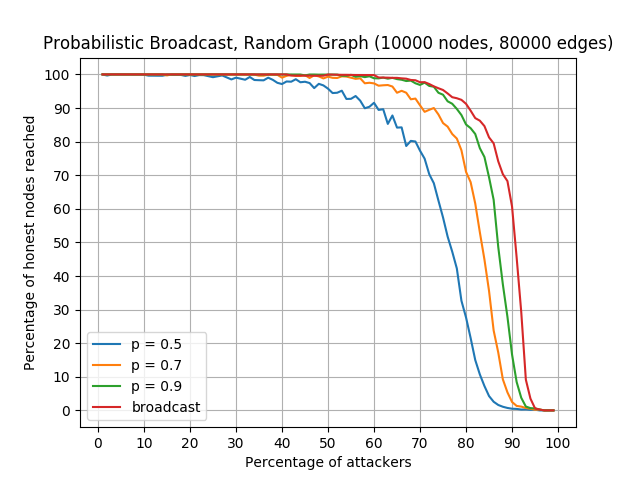}
  \caption{Coverage (Percentage of honest nodes reached) -- Dissemination: Probabilistic Broadcast, Overlay Topology: Random Graph with 80000 edges.}
  \label{random-coverage-probabilistic-80000}  
\end{figure}
\begin{figure}[t]
  \centering
\includegraphics[width=0.5\textwidth]{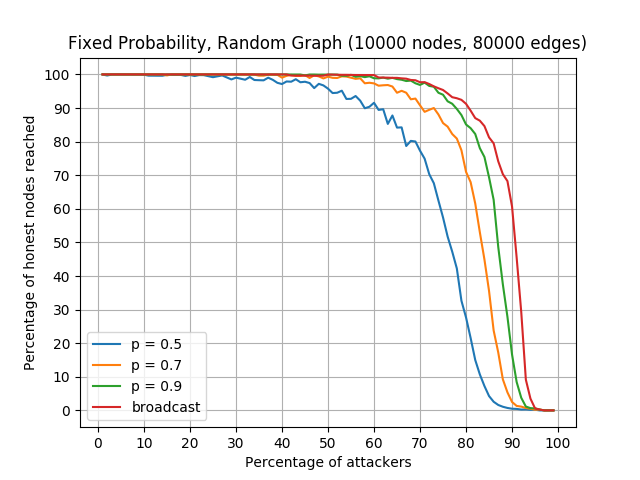}
  \caption{Coverage (Percentage of honest nodes reached) -- Dissemination: Fixed Probability, Overlay Topology: Random Graph with 80000 edges.}
  \label{random-coverage-fixed-80000}
\end{figure}

Finally, we used LUNES-Blockchain to study the effect of the network topology on the resilience to the Sybil attacks. More specifically, we repeated all the previously described tests but using small-world graphs instead of random graphs. The number of nodes and edges has been maintained equivalent. The results are not reported in this paper for conciseness but show that in this specific case the network topology has no relevant effects, in fact the obtained results are almost identical.

\section{Performance Evaluation of LUNES-Blockchain}\label{sec:simresults}

\begin{table*}[ht]
  \centering
  \caption{Execution time of the dissemination protocol (i.e.~LUNES) when varying on the number of LPs.}
\begin{tabular}{|c|c|c|c|c|c|c|}
\toprule
\textbf{Number of Edges }& \textbf{1 LP}   & \textbf{2 LP} & \textbf{3 LP} & \textbf{4 LP} & \textbf{5 LP} & \textbf{6 LP} \\
\midrule
\textit{500 nodes, 1500 edges}& 30s & 29s & 28s & 26s & 25s & 22s \\
\textit{500 nodes, 2000 edges} & 41s & 41s & 39s & 35s & 32s & 30s  \\
\textit{500 nodes, 2500 edges} & 53s & 51s & 47s & 45s & 41s & 39s  \\
\textit{1000 nodes, 4000 edges} & 331s & 209s & 184s &146s & 138s & 126s\\
\bottomrule
\end{tabular}
\label{LPtable}
\end{table*}

In this section, we analyze the runtime performance of the LUNES-Blockchain virtual environment. More specifically, we investigate how much the duration of the execution is dependent on the graph size and on the level of parallelization. We report this kind of analysis since we think that virtual environments should not only be able to provide correct and precise results but also to operate, in some specific cases, so fast that they can be used to operate real-world systems. For example, in support of real-time what-if analysis. All the tests reported in this section were executed on a PC equipped with an Intel Core i5 processor (8th generation) with 8 GBs of RAM running GNU/Linux Fedora 33 (kernel 5.8.17-300.fc33.x86\_64).

First of all, we have measured the amount of time required for completing the execution of the 51\% tests on a network with $500$ nodes and $2000$ edges. The sequential execution (i.e.~LUNES-Blockchain running on a single CPU core) required on average 19 minutes and 30 seconds. As expected, by doubling the number of edges the simulation required more than double the time, precisely 40m40s, while it required 46m09s to run with 1000 nodes and 4000 edges. When processing the data, we noticed slight variations among different executions, in the range of one minute, that are due to the random behavior of the gossip protocols implemented for the dissemination phase.

Analogously, selfish mining tests required, on average, 14m30s on a network with $500$ nodes and $2000$ edges, 33m17s on a network with double the edges and 36m21s on a graph with $1000$ nodes and $4000$ edges. It is interesting to note that the number of links affects the execution times more than the number of nodes. This can be easily explained by analyzing the behavior of the dissemination protocols since every increase in the number of edges has a huge impact on the number of messages that need to be disseminated in the network. This aspect has been already investigated in \cite{gda-jpdc-2017}.

As concerns the lower-level of the simulation (up to the implementation of the dissemination protocols), adopting a parallel approach (i.e.~multiple execution units are used for running the simulation) can speed up the execution of the LUNES-Blockchain component that implements this part of the simulation model (i.e.~LUNES).} Table~\ref{LPtable} reports the execution times, obtained using the Fixed Probability algorithm over a $500$ nodes graph with a varying number of edges of the network and using different configurations of LPs (i.e.~CPU cores used concurrently.

The data reported in the table suggest that even if several LPs are needed to achieve a significant speedup, it is possible to get a performance gain by using parallel execution. For example, with $5$ LPs it is possible to achieve a 20\% reduction of the execution time of this part of the simulator. 

We think that future systems, that will be based on larger networks and even more intensive communication patterns, will be able to further benefit of simulations following a PADS approach.

\section{Conclusions}\label{sec:conc}
In this paper, we have introduced a virtual environment built on top of an agent-based simulator called LUNES-Blockchain, which allows us to simulate the blockchain behaviour even in the presence of attacks. The virtual environment has been then used to verify the outcomes of the attacks in presence of different blockchain setups. Selfish mining and 51\% attack simulations showed that, as predictable, the feasibility of the attacks is directly proportional to the hash-rate owned by the malicious node. However, it was demonstrated that selfish mining is not an efficient strategy for the attacker if this one owns less than a half of the overall computing power. As concerns the 51\% attack, the most interesting consideration is that when the hash-rate $h\%$ of the attacker is $h>30\%$, the attacker is constantly able to mine a greater percentage of blocks, with respect to the percentage of computational power. On the other hand, denial of service attacks have a considerably greater success in the least connected  networks, because there is a greater probability that a node is connected to only malicious nodes. Finally, the simulation results showed that the gossip protocols used to spread information also have an impact on the resilience level, in particular an enhanced version of Dandelion that implements a fail-safe mechanism (in this paper referred as Dandelion+-) is able to fix the issues detected in the original version of Dandelion.
In general, LUNES-Blockchain demonstrated that virtual environments based on simulation can be used to study some relevant security aspects of complex systems such as blockchains. Moreover, when scalability is an issue, a parallel/distributed approach can give a limited but still relevant performance boost with respect to traditional methodologies based on sequential simulation.

\bibliographystyle{spmpsci}
\bibliography{biblio}

\end{document}